\documentclass[aps,prb,amsmath,amssymb,floatfix,twocolumn,superscriptaddress]{revtex4-1}
\usepackage{times}
\pdfoutput=1
\usepackage{bbold}
\usepackage{graphicx}

\begin{document}

\title{Pairing in  half-filled Landau level}
\author{Zhiqiang Wang}
\affiliation{Department of Physics and Astronomy, University of
California Los Angeles, Los Angeles, California 90095-1547, USA}
\author{Ipsita Mandal}
\affiliation{Perimeter Institute for Theoretical Physics, 31 Caroline St N, Waterloo, ON N2L 2Y5, Canada}
\author{Suk Bum Chung}
\affiliation{Center for Correlated Electron Systems, Institute for Basic Science (IBS), Seoul National University, Seoul 151-747, Korea} 
\affiliation
{Department of Physics and Astronomy, Seoul National University, Seoul 151-747, Korea }
\author{Sudip Chakravarty}
\affiliation{Department of Physics and Astronomy, University of
California Los Angeles, Los Angeles, California 90095-1547, USA}

\date{\today}

\begin{abstract}
  Pairing  of composite fermions in  half-filled Landau level state is reexamined by  solving the BCS gap equation with full frequency dependent current-current  interactions. Our results show that there can be a \emph{continuous} transition from the Halperin-Lee-Read state to a chiral odd angular momentum  Cooper pair state for  short-range contact interaction. This is at odds with the previously established conclusion of  first order pairing transition, in which the low frequency  effective interaction was  assumed for the entire frequency range. We find that even if the low frequency effective interaction is repulsive, it is compensated by the high frequency regime, which is attractive.   We  construct the phase diagrams and show that $\ell=1$ angular momentum channel is quite different from  higher angular momenta $\ell\ge 3$. Remarkably,  the full frequency dependent analysis applied to the bilayer Hall system with a total filling fraction $\nu=\frac{1}{2}+\frac{1}{2}$ is  quantitatively changed from the previously established results but not qualitatively.
\end{abstract}


\maketitle

\section{Introduction}
The concept of composite fermions was introduced to understand quantum Hall states.~\cite{Jain1989,*Jain2007} Using this concept, Halperin, Lee and Reed~\cite{Halperin1993} (HLR) further developed the Composite Fermi Liquid (CFL) theory to understand the gapless single layer half-filled Landau level problem. In the mean field approximation, this theory predicts a compressible ``metal" with a sharp Fermi surface, which has received some support from  experiments.\cite{Du1993,Jiang1989,Willett1990,Goldman1994,Kang1993,Willett1997} However, in such a gapless system, the emergent Chern-Simons (CS) gauge field fluctuations can play an important role. In fact, these fluctuations can mediate an attractive density-current interaction between the fluxes attached to a composite fermion and the current associated with another composite fermion. As was shown by Greiter, Wen and Wilczek~\cite{Greiter1992} (GWW), the mean field Fermi surface is always unstable to the formation of Cooper pairs in odd angular momentum channels. Thus  at low temperatures, the system  ends up in a superconducting state with an order parameter that is most likely a chiral
$p$-wave.

However, Bonesteel\cite{Bonesteel1999} showed that if we go beyond GWW analysis and consider the random phase approximation (RPA)~\cite{Halperin1993} corrections, there will be an induced current-current interaction mediated by the transverse CS gauge field. The current-current interaction is repulsive and divergent in the small Matsubara frequency limit for all angular momentum channels. Using this small frequency limit for the entire range of Matsubara frequencies, and assuming the gap to be  weakly dependent on  Matsubara frequencies, the zero temperature ($T=0$) BCS gap equation was solved analytically. The conclusion was that with a bare short-range contact interaction, the pair-breaking current-current interaction always dominates over other interactions when the BCS gap is small. Therefore there is no GWW pairing instablity. However the gap equation was found to have  have a solution when the gap was finite. Thus the zero temperature pairing transition was conjectured to be   first order for  short-range interaction, while for a long-range Coulomb interaction, a continuous pairing transition was considered to be  a possibility because of weaker gauge field fluctuations.

The above analysis,  based on the small frequency limit of the current-current interaction needs to be reexamined. A recent work~\cite{Chung2013} of our group involving non-relativistic fermions coupled to a transverse gauge field suggests that although the current-current interaction is repulsive and singular in the small frequency limit, at higher frequencies it can be attractive for angular momentum channels  $\ell\ge2$. This attractive part can outweigh the repulsive part, thus changing the solution to the BCS equation completely. Therefore it is crucial to consider the full Matsubara frequency dependence of the current-current interaction and solve the BCS gap equation in a self-consistent manner.

In addition to the single layer case, we also consider the full frequency dependent analysis of the double layer Hall system with a total filling fraction $\nu=\frac{1}{2}+\frac{1}{2}$. This system is decoupled into two separate composite fermion metals with $\nu=\frac{1}{2}$ in each layer if there are no disorder or inter-layer tunneling. If we include the most singular interaction, namely the inter-layer current-current interaction, the interaction can be attractive or repulsive depending on whether it is mediated by the out-of-phase or the in-phase mode of the CS gauge field fluctuations.~\cite{Bonesteel1993} The competition between these two will determine the final fate of the double layer system.  Bonesteel \emph{et al}~\cite{Bonesteel1996} showed that this interaction always drives the system into a inter-layer paired state for any large separation $d$ between the two layers. However, this conclusion was also obtained  from the previously mentioned small frequency analysis. Therefore, for similar reasons, it must be reexamined if only to put it on a firmer basis.

The major results of our paper are as follows: (1) For the single layer system with either short-range contact interaction or long-range Coulomb interaction, there can be a \emph{continuous} transition from the HLR state to a chiral odd $\ell$-wave Cooper pair state. (2) For the double layer system, there is always a non-zero pairing between inter-layer composite fermions if the inter-layer spacing $d$ is much greater than the magnetic length  $\ell_B$. Thus, the small frequency analysis~\cite{Bonesteel1996} does not qualitatively differ from the full frequency dependent analysis, except for significant quantitative differences.

The structure of the paper is as follows: in Sec. \ref{single-lyr formalism} we derive the BCS gap equation for the $\nu=\frac{1}{2}$ single layer system; in Sec. \ref{single-lyr results} we present our numerical results for the single layer system with short-range contact interaction and also briefly discuss the results for the long-range Coulomb interaction; in Sec. \ref{bi-layer formalism} we write down the BCS gap equation for the double layer system and present our numerical results; in Sec. \ref{conclusion} we summarize our conclusion and provide some further discussions.

\section{\label{single-lyr formalism}The Effective action and the BCS gap equation for a single layer system}
Consider a two-dimensional ($2D$) electron gas with a perpendicular magnetic field $B$ at a filling fraction $\nu=1/\tilde{\phi}$. For half-filling $\tilde{\phi}=2$. In the CFL picture, $\tilde{\phi}$, the emergent flux is  described by the CS gauge fields $(a_0,\mathbf{a})$  attached to an electron to form a composite fermion. To describe this flux attachment, we need to add a CS term to the free electron action.  Thus, without interactions, the total Euclidian Lagrangian density is given by~\cite{Halperin1993} $\mathcal{L}  = \mathcal{L}_{0}+\mathcal{L}_{CS}$ with ($\hbar= c=e=1$)
\begin{align}
 & \mathcal{L}_{0} =  \psi^*(\partial_\tau - a_0-\mu)\psi -\frac{1}{2m^{*}}\psi^*(\partial_i - i \mathbf{a}_i + i \mathbf{A}_i)^2  \psi \,, \\
 & \mathcal{L}_{CS}   =   \frac{a_0}{2\pi \tilde{\phi}} \, \epsilon^{i j}\partial_i a_j \,,
\end{align}
where $\psi$ is the composite fermion field, $m^{*}$ is the composite fermion effective mass, and $\mathbf{\nabla}\times\mathbf{A}=B \hat{\mathbf{z}}$, the physical magnetic field. At the mean field level, $ \epsilon^{i j}\partial_i \overline{a}_j= 2\pi \tilde{\phi} <\psi^*\psi> = B \hat{\mathbf{z}}$, $B$ being the applied magnetic field. Then the CS gauge field exactly cancels  the external magnetic field. Therefore the system is a composite fermion metal with the Fermi wave vector $k_F=\sqrt{\frac{2}{\tilde{\phi}}}\frac{1}{\ell_B}$, where $\ell_B$ is the magnetic length. At half filling, $k_F=\frac{1}{\ell_B}$.

Beyond this mean field approximation, there will be fluctuations of the composite fermion density, which also implies fluctuations of the CS gauge fields. These gauge field fluctuations can in turn couple to the composite fermion currents, mediating a density-current interaction. Such an interaction is attractive in odd angular momentum channels, leading to the GWW instability.
In addition, we have additional four fermion interaction term in the Lagrangian density,
\begin{equation}
 \mathcal{L}_{int} =\frac{1}{2}\int d^2 \mathbf{x}^{\prime}  \psi^*(\tau,\mathbf{x}) \, \psi(\tau,\mathbf{x})
 v (r) \, \psi^*(\tau,\mathbf{x}^{\prime}) \, \psi(\tau,\mathbf{x}^{\prime}) ,
\end{equation}
where $r=|\mathbf{x}-\mathbf{x}^{\prime}|$ and the interaction is:  $v(r)= \frac{e^2}{\epsilon r}$ for a Coulomb  interaction, $\epsilon$ being the dielectric constant, while $v(r)\propto  \delta(r)$ for a short-range contact interaction.

Redefining $(a_0,\mathbf{a})$ as  deviations from their mean field values, choosing the Coulomb gauge\cite{Halperin1993}, and using the constraint $\psi^*\psi=(1/2\pi \tilde{\phi}) \, \epsilon^{ij}\partial_i a_j$, we can rewrite the total action in the momentum space as
\begin{widetext}
\begin{align}
   S & =S_0+S_{CS} \,, \\
   S_0 & =\int d\tau \, d^2 \mathbf{x}
\left [ \psi^*(\partial_\tau -i a_0) \psi
-\frac{1}{2m^{*}}\psi^*(\partial_i - i \mathbf{a}_i)^2 \,\psi- \mu \, \psi^* \psi
\right ] \,, \\
   S_{CS} &= \int d\tau \, d^2\mathbf{x}
\left ( \mathcal{L}_{CS}+\mathcal{L}_{int} \right )
=\frac{1}{2 \beta}\sum_{ n, \mu,\nu}\int \frac{d^2 \mathbf{q}}{(2\pi)^2}
\, a_\mu^*(i\omega_n,\mathbf{q})
\, {\mathcal{D}_{\mu,\nu}^0}^{-1}(i\omega_n,\mathbf{q})
\, a_\nu(i\omega_n,\mathbf{q}) \,.
\end{align}
\end{widetext}
Here $\mu,\nu=\lbrace 0,1 \rbrace $, $\omega_n =2 n \pi/ \beta $ is a bosonic Matsubara
frequency, and $a_0,a_1$ are the time and transverse components of the CS gauge fields with the convention $a_1(i\omega_n,\mathbf{q})=\hat{\mathbf{z}}\cdot (\mathbf{q}\times \mathbf{a})$. The CS gauge field action term $S_{CS}$ defines the bare CS gauge field propagator to be
\begin{equation}
 \mathcal{D}^0(i\omega_n,\mathbf{q})
 =\left( \begin{array}{cc}
v(q) &    i\frac{2\pi \tilde{\phi}}{q}  \\
-i \frac{2\pi \tilde{\phi}}{q}                                &    0
\end{array} \right) \,,
\end{equation}
where $v(q)$ is the bare density-density interaction expressed in the momentum space. Hence, $v(q)=2\pi e^2/\epsilon q $ for  long-range Coulomb interaction, and $v(q)=const.$ for  short-range contact interaction. After RPA correction, the inverse gauge field propagator is given by
\begin{equation}
[\mathcal{D}(i\omega_n,\mathbf{q})]^{-1}  = [{\mathcal{D}^{0}}(i\omega_n,\mathbf{q})]^{-1}+\mathcal{K}^{0}(i\omega_n,\mathbf{q}) ,
\end{equation}
where $\mathcal{K}^{0}(i \omega_n,\mathbf{q})$ is the electromagnetic response function of the non-interacting fermions in $2D$ and has diagonal components only. In the limit $q<2k_F$ and $|\omega_n|\ll\frac{k_F \, q}{m}$, we $\mathcal{K}^{0}_{00}=N(\epsilon_F)=\frac{m^{*}} {2\pi}$ and $\mathcal{K}^{0}_{11}=-\chi_d \, q^2-\frac{k_F|\omega_n|}{2\pi q}$, where $\chi_d=\frac{1}{12\pi m^{*}}$ is the free fermion diamagnetic susceptibility. Inverting the matrix  $\mathcal{D}^{-1}$, the RPA corrected gauge field propagator is~\cite{Halperin1993}
\begin{widetext}
\begin{align}\label{eq:RPAgaugepropagator}
 & \mathcal{D}(i\omega_n,\mathbf{q})  =\frac{1}{\mathcal{K}_{00}^{0}
\left [ \mathcal{K}_{11}^{0}
- \frac{q^2 \, v(q)}{(2\pi \tilde{\phi})^2} \right ]
-\left (\frac{q}{2\pi \tilde{\phi}} \right )^2} \left( \begin{array}{cc}
\mathcal{K}^{0}_{11}
- \frac{q^2 \, v(q)}{(2\pi \tilde{\phi})^2}  &  -i\frac{q}{2\pi \tilde{\phi}}  \\
i \frac{q}{2\pi \tilde{\phi}}                                             & \mathcal{K}^{0}_{00}
\end{array} \right).
\end{align}
\end{widetext}
Therefore, the RPA correction not only renomalizes the density-density and density-current interactions, but also generates a new current-current interaction, which turns out to be repulsive and divergent at small frequencies but attractive at high frequencies.

Let $\phi_1$ and $\phi_2$ be the real and the imaginary parts of the anomalous self-energy respectively, and $Z(i\omega_n,{\mathbf{k}})$ denote the mass renormalization. For our  single layer problem we will ignore the mass renormalization equation by simply taking $Z(i\omega_n,{\mathbf{k}})\simeq 1$. This can be safely done in the pairing state if the superconductivity forms at an energy scale higher than the characteristic energy scale of the onset of the non-Fermi liquid behaviour.~\cite{Son1999} Therefore the complex anomalous self energy $\phi(i\omega_n,{\mathbf{k}})=\phi_1(i\omega_n,{\mathbf{k}})+i\phi_2(i\omega_n,{\mathbf{k}})$ is simply the gap function $\Delta(i\omega_n,{\mathbf{k}})$

Written out explicitly in terms of  $\Delta(i\omega_n,\mathbf{k})$, the anomalous self-energy equation in Matsubara frequency is~\cite{Schrieffer1999}
\begin{widetext}
\begin{gather}\label{eq:self-energy-2}
  \Delta(i\omega_n,{\mathbf{k}})=-\frac{1}{\beta}\sum_m \int\frac{d^2{\mathbf{k}}^{\prime}}{(2\pi)^2}\frac{\Delta(i\omega_m,{\mathbf{k}}^{\prime})}{|\omega_m |^2+\bar{\epsilon}_{\mathbf{k}^{\prime}}^2+|\Delta(i\omega_m,{\mathbf{k}}^{\prime})|^2}
\, V_{\rm{eff}}(i\omega_m -i \omega_n;{\mathbf{k}}^{\prime},{\mathbf{k}}) \,,
\end{gather}
as shown diagrammatically in Fig.~\ref{fig:anomalous-self-energy}. Here $\bar{\epsilon}_{\mathbf{k}}=\epsilon_{\mathbf{k}}-\mu$ is the reduced kinetic energy.
\begin{figure}
  \includegraphics[scale=0.5]{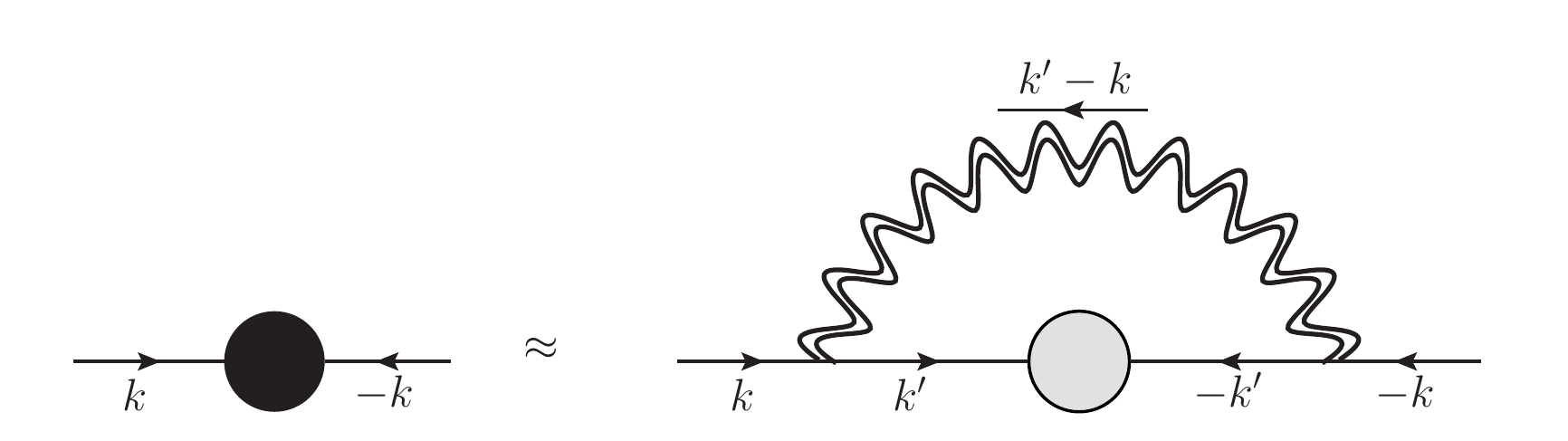}
  \caption{Diagrammatic representation of the Dyson equation for the off-diagonal components of the anomalous self-energy $\Delta(i\omega_n,\mathbf{k})$. The black blob denotes the proper anomalous self-energy, and the gray blob (along with the associated fermion lines) denotes the corresponding component of the exact propagator of the fermions. The wiggly line is the RPA corrected gauge field propagator. For compactness, we have used the Euclidean vector notation: $k \equiv (k_0, \mathbf{k})=(\omega_n, \mathbf{k})$.}
 \label{fig:anomalous-self-energy}
\end{figure}
\end{widetext}

Now let us write $\Delta(i\omega_n,{\mathbf{k}})=\sum_{\ell} \Delta_{\ell} (i\omega_n, {\mathbf{k}})   \, e^{i\ell \theta_{\mathbf{k}}}$, where only odd $\ell$ angular momentum channels will be considered since the $\nu=1/2$ single layer system is spin-polarized. Being a nonlinear integral equation involving the gap, we cannot strictly speaking decouple the various angular momentum channels. However, we shall
assume approximate decoupling. Such decoupled gap equations for different
channels can be considered as local minima of the free energy. We will use the Fermi surface approximation $|{\mathbf{k}}|=k_F$, ignoring the dependence on the magnitude of ${\mathbf{k}}$, so that $\Delta_{\ell}(i \omega_n, {\mathbf{k}}) \simeq \Delta_{\ell}(i\omega_n)$. We will also consider the zero temperature limit when the Matsubara frequencies become continuous: $\omega_n \rightarrow \omega$. For the $\ell$-wave channel, we get

\begin{widetext}
\begin{gather}
  \Delta_{\ell}(i\omega)=-\int\frac{d^2 \mathbf{k}^{\prime}}{(2\pi)^2}\frac{d\omega^{\prime}}{2\pi}
\frac{ \Delta_{\ell}(i\omega^{\prime})   \,  e^{i\ell \theta}   }
{|\omega^{\prime} |^2+\bar{\epsilon}_{\mathbf{k}^{\prime}}^2+|\Delta_{\ell}(i\omega^{\prime})|^2}
\, V_{\rm{eff}}(i\omega^{\prime} -i \omega;{\mathbf{k}}^{\prime},{\mathbf{k}}) \Big|_{|\mathbf{k}|, |\mathbf{k^{\prime}}|=k_F}  \,,
\end{gather}
where $\theta$ is defined as $\hat{\mathbf{z}}\cdot (\hat{\mathbf{k}}\times \hat{\mathbf{k}}^{\prime})=\sin\theta$.
\end{widetext}
 At the Fermi surface, $\int \frac{d^2 \mathbf{k}^{\prime}}{(2\pi)^2}\approx N(\epsilon_F)\int_{-\infty}^{\infty} d\bar{\epsilon}_{\mathbf{k^{\prime}}} \int \frac{d\theta}{2\pi}$, where $N(\epsilon_F)=\frac{m^{*}}{2\pi}$ is the $2D$ density of states of spin-polarized fermions at the Fermi energy $\epsilon_F$. Performing the integration over $\bar{\epsilon}_{\mathbf{k^{\prime}}}$ by the contour integral method, the BCS gap equation takes the form:
\begin{widetext}
\begin{gather}\label{eq:BCS-gap}
   \Delta_{\ell}(i\omega)=\int d \omega^{\prime}\frac{ \Delta_{\ell}(i\omega^{\prime})}{2\sqrt{|\omega^{\prime}|^2+|\Delta_{\ell}(\omega^{\prime})|^2}}\widetilde{V}_{\rm{eff},\ell}(|\omega - \omega^{\prime}|) \,,\\
  \widetilde{V}_{\rm{eff},\ell}(|\omega -\omega^{\prime}|)=
 -\frac{m^{*}}{2\pi}\int \frac{d\theta}{2\pi}  \,
e^{i\ell \theta}
\, V_{\rm{eff}}(i\omega^{\prime}-i \omega;{\mathbf{k}}^{\prime},{\mathbf{k}}) \Big|_{|\mathbf{k}|, |\mathbf{k^{\prime}}|=k_F} \,.
\end{gather}
\end{widetext}
The Fermi surface  approximation is good in the BCS  case, because the Debye frequency $\omega_D$ is much smaller than the typical Fermi energy. But in the present problem, there is no similar small energy scale. There is, however,  a rough high energy scale  $\omega_{0}\sim\frac{e^2}{\epsilon \ell_B}$, where $\ell_B$ is the magnetic length. On a length scale smaller than $\ell_B$, the concept of composite fermion is not well-defined. But this energy scale is not small compared with $\epsilon_F$; in fact, $\frac{\omega_0}{\epsilon_F}=\frac{2m^{*} e^2 \, \ell_B}{\epsilon} \simeq \frac{20}{3}$.~\cite{Bonesteel1999, Halperin1993} It has the same order of magnitude as the Fermi energy of the free composite fermions. We assume that  our Fermi surface approximation to be qualitatively correct.
To get the explicit form of $V_{\rm{eff}}$, we  need to multiply $\mathcal{D}_{\mu , \nu}$ by the appropriate vertex factors, as shown in the Feynman diagram in Fig.~\ref{fig:anomalous-self-energy}. As we mentioned before, there will be density-density (associated with $\mathcal{D}_{00}$), density-current (involving $\mathcal{D}_{01}$ and $\mathcal{D}_{10}$), and current-current interactions (involving $\mathcal{D}_{11}$).

Correspondingly $\widetilde{V}_{\rm{eff},\ell}(i\omega)$, which characterizes the interaction, can be separated into three pieces: the density-density interaction term $\lambda_{\ell,00}(i\omega)$, the density-current interaction term $\lambda_{\ell,10}(i\omega)$, and the current-current interaction term $\gamma_{\ell}(i\omega)$:
 \begin{gather}
  \widetilde{V}_{\rm{eff},\ell}(|\omega|) = \lambda_{\ell,00}(i\omega) + \lambda_{\ell,10}(i\omega)- \gamma_{\ell}(i\omega)
 \end{gather}
 with $\lambda_{\ell,00},\lambda_{\ell,10}$ and $\gamma_{\ell}$ are defined by
\begin{widetext}
\begin{align}
  &  \lambda_{\ell,00}(i\omega)\equiv -\frac{m^{*}}{2\pi}\int_{0}^{2\pi}\frac{d\theta}{2\pi}
\,e^{i\ell \theta} \; \mathcal{D}_{00}(i\omega,|{\mathbf{k}}^{\prime}-{\mathbf{k}}|)
\Big|_{|\mathbf{k}|, |\mathbf{k^{\prime}}|=k_F} \,. \\
&  \lambda_{\ell,10}(i\omega)\equiv -\frac{m^{*}}{2\pi}\int_{0}^{2\pi}\frac{d\theta}{2\pi}
\,e^{i\ell \theta} \;
 \left \{ 2\frac{ \,  \hat{\mathbf{z}} \cdot ({\mathbf{k}}^{\prime}\times{\mathbf{k}})}{m^{*}|{\mathbf{k}}^{\prime}-{\mathbf{k}}|}\mathcal{D}_{01}(i\omega,|{\mathbf{k}}^{\prime}-{\mathbf{k}}|) \right \}
\Big|_{|\mathbf{k}|, |\mathbf{k^{\prime}}|=k_F} \,.  \\
  & \gamma_{\ell}(i\omega)\equiv \frac{m^{*}}{2\pi}\int_{0}^{2\pi}\frac{d\theta}{2\pi}
\,e^{i\ell \theta} \;
\left \{\frac{|\hat{\mathbf{z}}\cdot({\mathbf{k}}^{\prime}\times{\mathbf{k}})|^2}{{m^{*}}^2|{\mathbf{k}}^{\prime}-{\mathbf{k}}|^2}\mathcal{D}_{11}(i\omega,|{\mathbf{k}}^{\prime}-{\mathbf{k}}|) \right \}
\Big|_{|\mathbf{k}|, |\mathbf{k^{\prime}}|=k_F} \,. \label{eq: gamma-ell-def}
\end{align}
\end{widetext}
In the density-current interaction term $\lambda_{\ell,10}(i\omega)$, there is a pre-factor of $2$ inside the curly brackets. This is because the two off-diagonal CS gauge field propagators $\mathcal{D}_{01}$ and $\mathcal{D}_{10}$ contribute identically to the density-current interaction.

It can be shown that for the short-range interaction $v(q)=const$
\begin{align}
  \lambda_{\ell,00}(i\omega=0)& =0 \\
  \lambda_{\ell,10}(i\omega=0)& =\xi_{\ell} \; \mathrm{sgn}(\ell)
\end{align}
with $\xi_{\ell}$ some positive constant~\cite{Bonesteel1999}.  From the expression of $\lambda_{\ell,10}(i\omega=0)$ we see that it is attractive in the positive $\ell-$wave channel while repulsive in the negative $\ell-$wave channel, indicating the chirality of pairing if  superconductivity is stabilized. Notice that because of the Fermi surface approximation $\xi_{\ell}$ is independent of angular momentum channel $\ell$. However as this approximation is relaxed, in general it should pick up an $\ell$ dependence. In the following we are going to group the density-density and density-current interaction together and use $\xi_{\ell} = \lambda_{\ell,00}(i\omega=0)+\lambda_{\ell,10}(i\omega=0)$ (for $\ell >0$) as a generic coupling constant to characterize them.

The transverse component gauge field propagator is given by~\cite{Bonesteel1999}:
\begin{gather}
 \mathcal{D}_{11}(i\omega,q=|{\mathbf{k}}^{\prime}-{\mathbf{k}}|)  \simeq \frac{1}{\tilde{\chi}(q)q^2+\frac{k_F |\omega|}{2\pi q}}
\end{gather}
where $\tilde{\chi}(q)=v(q)/(2\pi\tilde{\phi})^2+(1+6/\tilde{\phi}^2)/(12\pi m^{*})$. With short-range interaction $v(q)\simeq v(0)=const$, $\tilde{\chi}(q)\simeq \tilde{\chi}(0)$ is a constant. Then $\gamma_{\ell}(i\omega)$ of Eq.~\eqref{eq: gamma-ell-def} can be expressed as:

\begin{equation}
\gamma_{\ell}(i \omega)  = \zeta_{\ell} \int_{0}^{2\pi} \frac{d\theta}{2\pi}
 \frac{ |\sin \frac{\theta}{2}| \, \cos^2 \frac{\theta}{2}\cos(\ell\theta)}{2
|\sin^3\frac{\theta}{2}|
+\zeta_{\ell}\frac{|\omega|}{4 \, \epsilon_F}} \,.         \label{eq:gamma-ell-short}
\end{equation}
with the dimensionless  constant $\zeta_{\ell}$ defined by
\begin{gather}
\zeta_{\ell}\equiv \frac{1}{4\,\pi \, m^{*}\, \tilde{\chi}(0)} \,
\end{gather}
Similarly to $\xi_{\ell}$, $\zeta_{\ell}$ will be also treated as a generic coupling
constant for the current-current interaction.

In the small frequency limit
\begin{gather}
\gamma_{\ell}(i\omega) \propto  (\frac{\epsilon_F}{|\omega|})^{1/3}
\end{gather}
with a frequency independent prefactor. Bonesteel~\cite{Bonesteel1999} used this small frequency expression for the whole frequency range and solved the BCS gap equation. However, according to our previous experience~\cite{Chung2013} the behavior of $\gamma_{\ell}(i\omega)$ at high frequency is qualitatively very different from that in the small frequency limit. In fact it changes sign at high frequencies. Therefore, the full frequency dependence of $\gamma_{\ell}(i\omega)$ will be very important for the final result. In other words, we will use Eq.~\ref{eq:gamma-ell-short} for $\gamma_{\ell}(i\omega)$.

\begin{figure}[ht]
  \centering
  \includegraphics[scale=0.7]{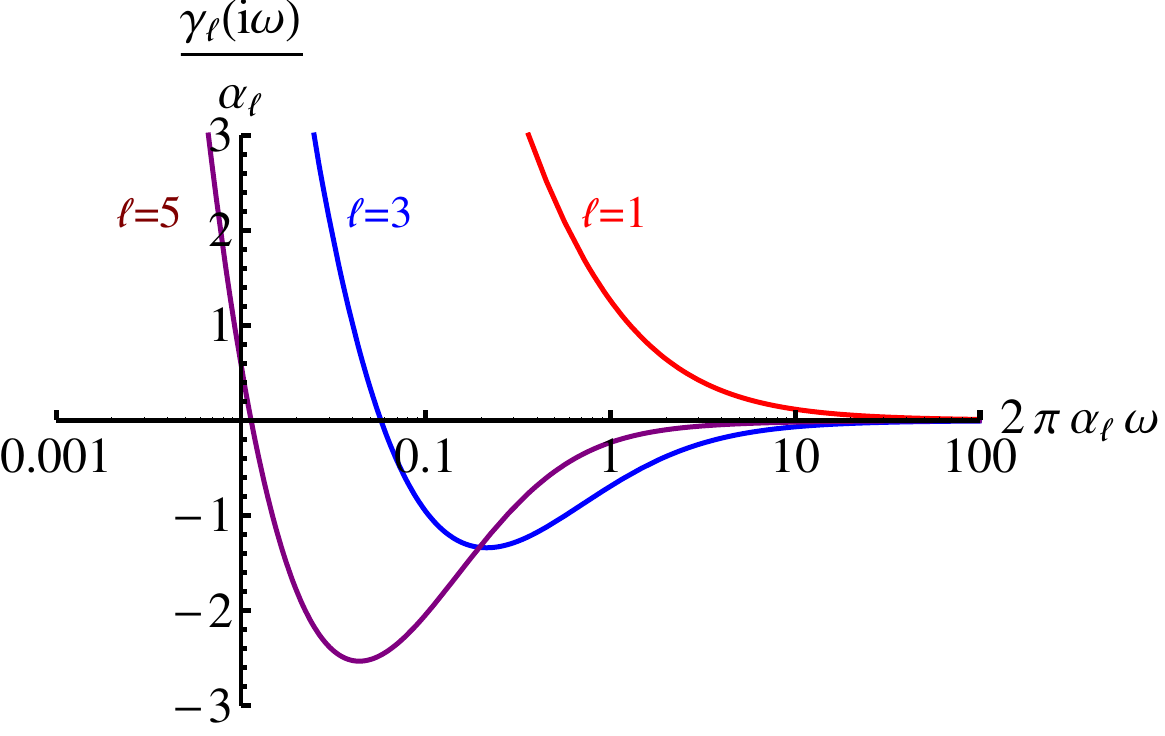}
  \caption{$\frac{\gamma_{\ell}(i\omega)}{\alpha_{\ell}}$ versus $2\pi \alpha_{\ell} \, \omega$ for the short-range interaction case, where $\alpha_{\ell}\equiv \frac{\zeta_{\ell}}{4\pi}>0$. The frequency $\omega$ has been expressed in units of $\epsilon_F$.}
  \label{fig:gamma-omega-1-3-5}
\end{figure}

Fig.~\ref{fig:gamma-omega-1-3-5} shows the dependence of $\gamma_{\ell}(i\omega)$ on frequency $\omega$. From these plots it is transparent that the current-current interaction term becomes attractive at high frequencies for angular momentum channel $\ell > 1$. Furthermore the attractive parts are considerable.

\begin{widetext}
In summary, the equation  to be solved is
\begin{equation}
\Delta_{\ell}(\omega) =\xi_{\ell} \; \int_{-\omega_{c1}}^{\omega_{c1}} d\omega^{\prime} \,
\frac{\Delta_{\ell}(\omega^{\prime})
}
{2\sqrt{{\omega^{\prime}}^2+|\Delta_{\ell}(\omega^{\prime})|^2}}\, - \int_{-\omega_{c2}}^{\omega_{c2}} d\omega^{\prime} \,
\frac{\Delta_{\ell}(\omega^{\prime})
}
{2\sqrt{{\omega^{\prime}}^2+|\Delta_{\ell}(\omega^{\prime})|^2}} \; \gamma_{\ell}(i\omega)
\end{equation}
\end{widetext}
where the full frequency dependence, Eq.~\ref{eq:gamma-ell-short} of $\gamma_{\ell}(i\omega)$, is used.

In the above gap equation we have set a frequency cut-off $\omega_{c1}$ for the  term involving  $\xi_{\ell}$. Using different values of $\omega_{c1}$ can in general change the non-universal critical  constants  $\xi_{\ell}$  and $\zeta_{\ell}$ of the phase transition. However, as far as the nature of the phase transition is concerned, which is what we are interested in, the conclusions obtained here will be independent of these specific values. For simplicity we will choose $\omega_{c1}=\epsilon_F$. As for the current-current interaction term, we are going to choose the frequency integration cutoff $\omega_{c2}$ to be large enough to include all significant contributions of $\gamma_{\ell}(i\omega)$. Typical value of this cutoff for the  displayed numerical results is $\omega_{c2}=10\, \epsilon_F$; in Ref.~\onlinecite{Bonesteel1999},  $\omega_{c2}$ was set to infinity.

\section{\label{single-lyr results}Numerical results for single layer system}

\begin{figure}[ht]
  \centering
  \includegraphics[scale=0.7]{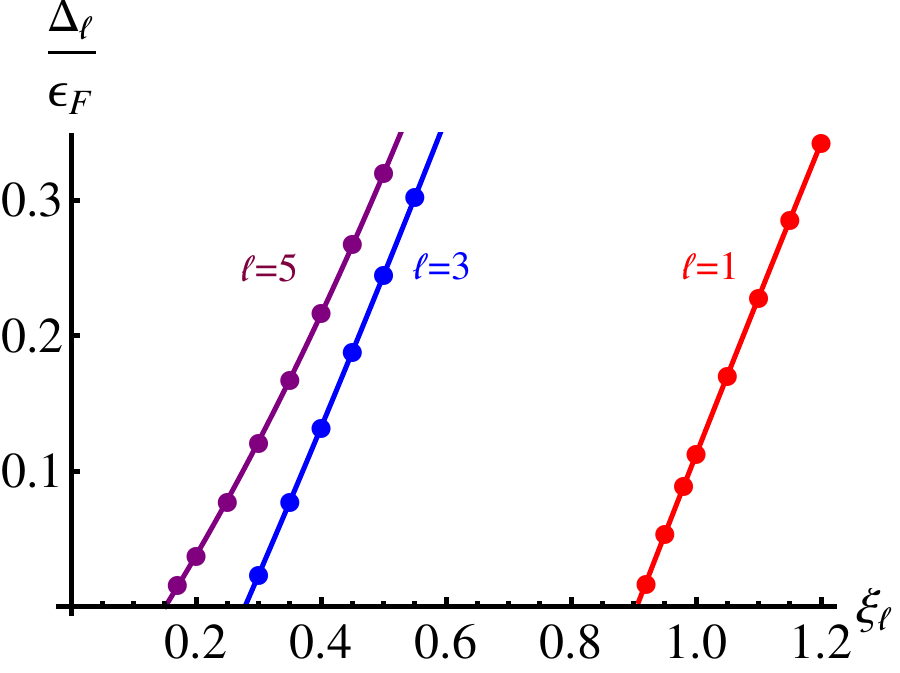}
  \caption{Zero frequency gap $\Delta_{\ell}$ versus $\xi_{\ell}$ for the short-range interaction, plotted for $\ell=1,3,5$. The other coupling constant $\zeta_{\ell}=1$.}
\label{fig:short-gap-xi-1-3-5}
\end{figure}
Fig.~\ref{fig:short-gap-xi-1-3-5} shows the zero frequency gap $\Delta_{\ell}\equiv \Delta_{\ell}(i\omega=0)$ versus $\xi_{\ell}$ for different $\ell$'s, with $\zeta_{\ell}=1$ (fixed). There are several noticeable features in this graph: (1) When $\xi_{\ell}$ is large enough, superconductivity exists. (2) The phase transition  is continuous for all odd values of $\ell$. (3) For $\ell=1$, pairing requires a  larger value of $\xi_{\ell}$, because in this case, $\gamma_{\ell}(i\omega)$ is repulsive for all frequencies and a larger attraction from $\xi_{\ell}$ is required to produce pairing, as is shown previously in Fig.~\ref{fig:gamma-omega-1-3-5}.

\begin{figure}[ht]
  \centering
  \includegraphics[scale=0.7]{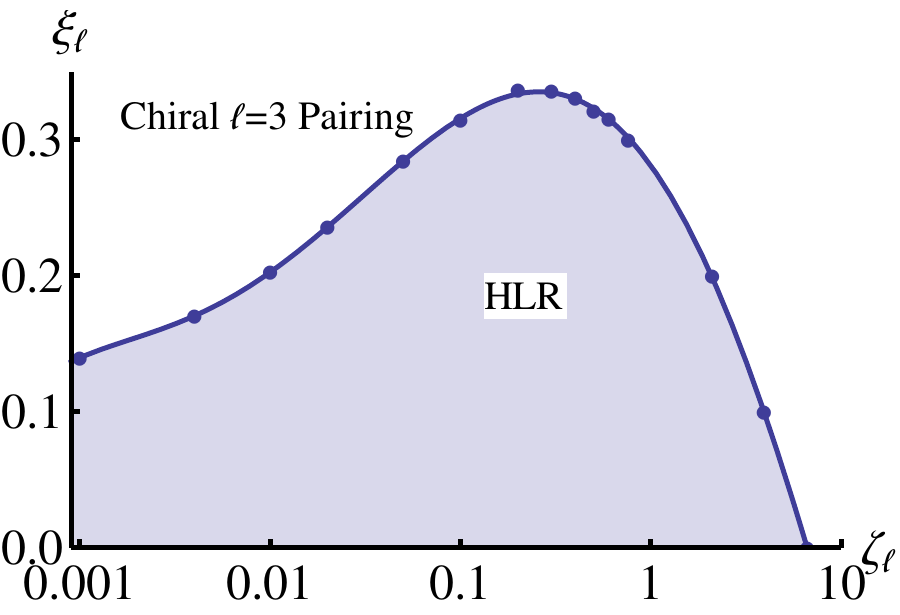}
  \caption{Phase diagram for $\ell=3$ in the $\xi_{\ell}$-$\zeta_{\ell}$ plane, for the case of short-range interaction. In the shaded region, we have an \emph{HLR} state. In the other regions, the system is unstable to a \emph{chiral $\ell=3$ pairing state}. The transition from the pairing state to the HLR state is \emph{continuous}. We note that when $\zeta_{\ell}$ is very small, the phase boundary curve should be extrapolated to the origin point. }
  \label{fig:ell-3-phasediag-short}
\end{figure}

We have solved the gap equation for different values of $\zeta_{\ell}$ to find the corresponding critical values of $\xi_{\ell}$ and constructed the phase diagram in the $\xi_{\ell}-\zeta_{\ell}$  space. The phase diagram for $\ell=3$  is shown in Fig.~\ref{fig:ell-3-phasediag-short}.
The generic  features for $\ell> 1$ are as follows:
\begin{enumerate}
\item   Pairing exists when $\zeta_{\ell}$ is large, since each individual term of the gap equation leads to pairing. The threshold value of $\xi_{\ell}$ for pairing increases as $\zeta_{\ell}$ decreases.

\item  When $\zeta_{\ell}$ is small and $\xi_{\ell}$ is not large enough, the current-current interaction term is pair breaking, while the density-current term is not powerful enough to overcome this effect. Hence HLR state is stable against pairing in this region.
\item  If $\zeta_{\ell}$ is very small, we can ignore the current-current interaction term. Then even small $\xi_{\ell}$ can give us a pairing state. This is because in the limit $\zeta_{\ell}\rightarrow 0$, even infinitesimal attraction, characterized by $\xi_{\ell}$, will produce a pairing state. This is why in Fig.~\ref{fig:ell-3-phasediag-short} when $\zeta_{\ell}$ is very small, the phase boundary could be extrapolated to the origin.
\end{enumerate}

\begin{figure}[ht]
  \centering
  \includegraphics[scale=0.7]{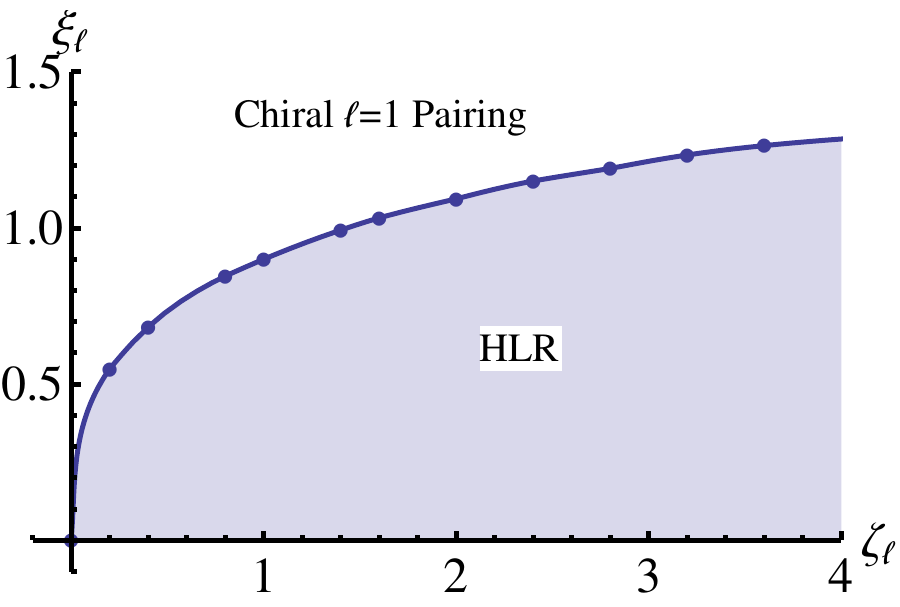}
  \caption{Phase diagram for $\ell=1$ in the $\xi_{\ell}$-$\zeta_{\ell}$ plane, for the case of short-range interaction. The phase transition across the phase boundary is continuous.}
  \label{fig:ell-1-phasediag-short}
\end{figure}

As the frequency dependence of $\gamma_{\ell}(i\omega)$ is similar for all $\ell\ge 3$, seen in Fig.~\ref{fig:gamma-omega-1-3-5} , the phase diagrams for different $\ell\ge 3$ should also be similar. The $\ell=1$ phase diagram can be quite different since the current-current interaction term for this channel is  repulsive over the entire frequency range. Only when the attractive term $\xi_{\ell}$ is large enough, overcoming  the repulsive current-current interaction, a chiral $\ell=1$ pairing state can exist. This is explicitly shown in the $\ell=1$ phase diagram in Fig.~\ref{fig:ell-1-phasediag-short}.

Although all the above calculations are done with short-range contact interaction, we have also checked the Coulomb interaction case. In this case, the attractive part from the current-current interaction at high frequency is very small. So the current-current interaction mostly serves as a pair breaking term. Therefore the conclusion is similar to what Bonesteel~\cite{Bonesteel1999} had in his paper: only when the density-current interaction dominates over other interactions will we have the GWW instability and the system be in a chiral odd $\ell$ pairing state; also the transition from the HLR state to the chiral pairing state is continuous. We should remind the reader that in reality the long-range interaction may take some different form, such as $1/q^x$, with $x\le 2$, some positive constant~\cite{Halperin1993, Nayak1994}. Although we did not do explicit calculations for these cases, we expect that the general conclusions will be the same as that obtained for the Coulomb interaction.


\section{\label{bi-layer formalism} BCS gap equation for the double layer system}
The full frequency dependent analysis can  also be applied to the double layered Landau level system with a total filling fraction $\nu=\frac{1}{2}+\frac{1}{2}$ without inter-layer tunneling. Originally Bonesteel \textit{et al}~\cite{Bonesteel1996} used full Eliashberg equations in their analysis. But it turns out that, as far as whether there is a BCS pairing state at zero temperature or not, the single BCS gap equation is enough. Since this will be a qualitative discussion, to simplify our numerical task, we are going to ignore the second Eliashberg equation; in other words set $Z\approx 1$, as in the single layer problem. The numerical solution of the coupled Eliashberg equations with the full frequency dependence is quite complex.

In the zero temperature limit, the BCS gap equation for $s$-wave pairing in the Matsubara frequency space is~\cite{Bonesteel1996}
\begin{gather}\label{eq:BCS-bilyr}
\Delta(\omega) = \int_{-\omega_c}^{\omega_{c}} d\omega^{\prime} \,
\frac{\Delta(\omega^{\prime})
}
{2 \sqrt{{\omega^{\prime}}^2+|\Delta(\omega^{\prime})|^2}}\, \tilde{V}_{\rm{eff}}(|\omega^{\prime}-\omega|)
\end{gather}
where the effective interaction $V_{\rm{eff}}$ has two terms
\begin{gather}
\tilde{V}_{\rm{eff}}(|\omega|)= \lambda^{(-)}(i\omega) - \lambda^{(+)}(i\omega)
\end{gather}
Here the two dimensionless coupling constants $\lambda^{(\pm)}(i\omega)$ are defined as the average of two inter-layer current-current interactions in the Cooper channel over the Fermi surface
\begin{gather} \label{eq:lambda-bilyr}
\lambda^{(\pm)}(i\omega)= \frac{m^{*}}{2\pi} \int_{0}^{2\pi}\frac{d\theta}{2\pi}
\,\left ( \frac{{\mathbf{k}}\times \hat{\mathbf{q}}}{m^{*}} \right )^2 \, \mathcal{D}^{\pm}(i\omega,q)\;
\end{gather}
with $\mathbf{q}={\mathbf{k}}-{\mathbf{k}}^{\prime}$, $ q = |\mathbf{q}|$. The scattering angle $\theta$ is defined via $\hat{\mathbf{z}}\cdot (\hat{\mathbf{k}}\times \hat{\mathbf{k}}^{\prime})=\sin \theta$, same as before.

Here the superscript $\pm$ means that the interaction is mediated either by the inter-layer in-phase (`$+$' sign) or the out-of-phase (`$-$' sign) modes of the gauge field fluctuations. And $\mathcal{D}^{\pm}(i\omega, q )$ are the two corresponding gauge field fluctuation propagators. Within RPA and in the limit that the inter-layer spacing $d\gg \ell_B$, they are given by
\begin{gather}
\mathcal{D}^{+}(i\omega,q)\simeq
\left ( \frac{e^2 \,q}{4\pi \epsilon}+\frac{|\omega|\, k_F}{4\pi q} \right )^{-1} \,,
\end{gather}
and
\begin{gather}
\mathcal{D}^{-}(i\omega,q)\simeq \left\{
      \begin{array}{lr}
      \left (\frac{e^2 \, d \, q^2}{4\pi \epsilon}+\frac{|\omega| \, k_F}{4\pi q} \right )^{-1} \,, & \text{for } q \lesssim d^{-1} \,,\\
       \left ( \frac{e^2 \,q}{4\pi \epsilon}+\frac{|\omega|\,k_F}{4\pi q} \right )^{-1}  \,, & \text{for } q \gtrsim d^{-1} \,.
      \end{array}
     \right .
\end{gather}
where $\epsilon$ is the dielectric constant. 
\begin{widetext}
Now we can substitute these expressions into our definitions of $\lambda^{(\pm)}(i \omega)$ in Eq.~\eqref{eq:lambda-bilyr}. Writing everything out in terms of the scattering angle $\theta$ and making the Fermi surface approximation yields (with $k_F=\ell_B^{-1}$ at half-filling)
\begin{gather}
 \lambda^{(+)}(i\omega)  = \int_{0}^{2\pi}\frac{d\theta}{2\pi} \frac{4\sin\theta}{\beta(2-2\cos\theta)+\frac{|\omega|}{\epsilon_F}}  \\
 \lambda^{(-)}(i\omega)  =
      [\int_{0}^{\theta_c}+\int_{2\pi-\theta_c}^{2\pi}]\frac{d\theta}{2\pi} \frac{4\sin\theta}{\beta \frac{d}{\ell_B}(2-2\cos\theta)^{3/2}+\frac{|\omega|}{\epsilon_F}}
     +  \int_{\theta_c}^{2\pi-\theta_c}\frac{d\theta}{2\pi} \frac{4\sin\theta}{\beta(2-2\cos\theta)+\frac{|\omega|}{\epsilon_F}}  \,,
\end{gather}
\end{widetext}
where $\theta_c=2 \arcsin (\frac{1}{2\; d/\ell_B})$. In the expression of $\lambda^{(-)}$ above, the first term comes from the propagator $\mathcal{D}^{-}(i\omega,q)$ with $q \lesssim d^{-1}$. We have also introduced the dimensionless paramter of $\beta \equiv \frac{e^2/(\epsilon \ell_B)}{\epsilon_F}=\frac{e^2m^{*}}{\epsilon k_F}$. Direct inspections show that the $\int_{\theta_c}^{2\pi-\theta_c}$ part of integral contribution is the same in both $\lambda^{(\pm)}$. Therefore they cancel out each other in the effective interaction $\widetilde{V}_{\rm{eff}}$. This cancellation is vital to the explanations to our final numerical results. After this cancellation $\widetilde{V}_{\rm{eff}}=\lambda^{(-)}-\lambda^{(+)}$ is simply
\begin{widetext}
\begin{align} \label{eq:bilyr-Veff}
\widetilde{V}_{\rm{eff}}(i\omega)=\int_{0}^{\theta_c}\frac{d\theta}{\pi} \;\big \{\frac{4\sin\theta}{\beta \frac{d}{\ell_B}(2-2\cos\theta)^{3/2}+\frac{|\omega|}{\epsilon_F}}\; - \frac{4\sin\theta}{\beta(2-2\cos\theta)+\frac{|\omega|}{\epsilon_F}}\big\}.
\end{align}
\end{widetext}
This gives the full-frequency dependent $\widetilde{V}_{\rm{eff}}(i\omega)$. In the original analysis of Ref.\onlinecite{Bonesteel1996}, the zero frequency limit $\omega\rightarrow 0$ expression of $\widetilde{V}_{\rm{eff}}(i\omega)$ is used. Since we are going to make a comparison between the solutions using these two different expressions of $\widetilde{V}_{\rm{eff}}(i\omega)$, we also give the small frequency limit expression of $\widetilde{V}_{\rm{eff}}(i\omega)$ here. Taking the $\omega \rightarrow 0$ limit of Eq.~\eqref{eq:bilyr-Veff} and keeping the most singular part only in each term we obtain
\begin{widetext}
\begin{align} \label{eq:bilyr-Veff-small}
\widetilde{V}_{\rm{eff}}(i\omega)  \simeq  \frac{8}{3\sqrt{3}} \; ( \frac{1}{\beta \; d/\ell_B})^{2/3} (\frac{\epsilon_F}{|\omega|})^{1/3} - \; \frac{2}{\pi} \frac{1}{\beta} \; \ln\frac{\epsilon_F}{|\omega|}
\end{align}
\end{widetext}
From this expression it is obvious that the out-of-phase attraction $\sim (\frac{\epsilon_F}{|\omega|})^{1/3}$ is more singular than, therefore dominates over, the in-phase repulsion $\ln \frac{\epsilon_F}{|\omega|}$ in the zero frequency limit. Hence using this $\widetilde{V}_{\rm{eff}}$ always gives us pairing. What we want to see here is whether including the full frequency dependence of $\widetilde{V}_{\rm{eff}}(i\omega)$ is going to change the conclusions or not.

The specific value of $d$ does not change our conclusion as long as it satisfies $d \gg \ell_B$, which is the domain of validity for all the above discussions. As for the frequency cutoff $\omega_c$ in the BCS equation~\eqref{eq:BCS-bilyr}, $\omega_c=10 \epsilon_F$ is taken when the full frequency dependent $\widetilde{V}_{\rm{eff}}$~\eqref{eq:bilyr-Veff} is used. This value should be large enough to include all significant high frequency contributions. When using the small frequency limit expression of $\widetilde{V}_{\rm{eff}}$~\eqref{eq:bilyr-Veff-small} we can still use a large frequency cutoff $\omega_c=10\epsilon_F$ for the first term involving $|\omega|^{-1/3}$, as it decays fast enough at high frequencies. However, the second term involving $\ln \frac{\epsilon_F}{|\omega|}$ does not converge at high frequencies. Therefore we will set $\omega_c=\epsilon_F$ for it.


\subsection*{\label{bi-layer results} Numerical results for double layer Hall system}

\begin{figure}[ht]
\centering
  \includegraphics[scale=0.76]{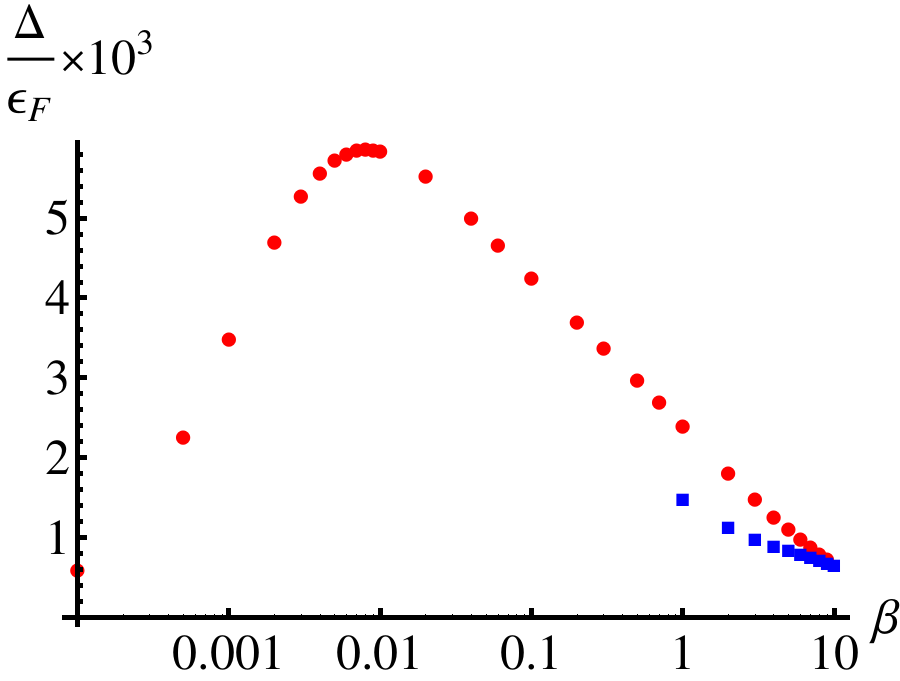}
  \caption{Pairing gap $\Delta$ for different values of $\beta \equiv \frac{e^2/(\epsilon \ell_B)}{\epsilon_F}$, in the $s$-wave channel. The circles denote the data obtained by using full frequency dependent $\widetilde{V}_{\rm{eff}}$, Eq.~\eqref{eq:bilyr-Veff}, while the squares denote the data obtained by using its small frequency limit expression, Eq.~\eqref{eq:bilyr-Veff-small}. Inter-layer distance $d  k_F=\frac{d}{\ell_B}=10$.}
  \label{fig:bilayer-gap-beta-comparison}
\end{figure}

Fig. \ref{fig:bilayer-gap-beta-comparison} shows the zero frequency gap $\Delta \equiv \Delta(i\omega=0)$ as a function of  $\beta $.  Clearly, we always get non-zero pairing, irrespective of whether the full frequency dependence is considered or not.
The explanation is as follows. As we mentioned before, the two gauge field propagators $\mathcal{D}^{\pm}(i\omega,q)$ differ from each other only when $q\lesssim d^{-1}$. This corresponds to a small frequency range $|\omega|\lesssim \omega_c$, with $\omega_c$ estimated by equating the Coulomb interaction energy term with the Landau damping term in the denominator of $\mathcal{D}^{+}(i\omega_c,q=d^{-1})$
\begin{gather}
\frac{e^2 \, q}{4\pi \epsilon} \simeq \frac{\omega_c \, k_F}{4\pi q} \Rightarrow \frac{\omega_c}{\epsilon_F} \simeq \frac{e^2/(\epsilon \, \ell_B)}{\epsilon_F} (q  \,\ell_B)^2 \simeq \beta \left ( \frac{\ell_B}{d} \right )^2 \ll 1 \,.
\end{gather}
Therefore, $\lambda^{(+)}(i\omega)$ is significantly different from $\lambda^{(-)}(i\omega)$ only when $|\omega|/\epsilon_F \ll 1$. This implies that $\widetilde{V}_{\rm{eff},\ell}(i\omega)$ is basically non-vanishing only in the range $|\omega|/\epsilon_F \lesssim \omega_c/\epsilon_F\ll 1$. Hence including the high frequency part of $\widetilde{V}_{\rm{eff}}(i\omega)$ does not qualitatively change the conclusion. Of course quantitatively there will be differences as we see in Fig.~\ref{fig:bilayer-gap-beta-comparison}. In fact this difference is bigger when $\beta$ is smaller. This is because $\beta$ comes into $\widetilde{V}_{\rm{eff}}(i\omega)$ in the form of $(\frac{1}{\beta})^{x}$ with $x=2/3, 1$ in the prefactors. Therefore when $\beta$ is smaller, the quantitative difference is enhanced. Notice that we only present our numerical gap data for $\beta \ge 1$ when using the small frequency limit expression of the effective interaction Eq.~\eqref{eq:bilyr-Veff-small}. This is because when $\beta \ll 1$, taking the frequency cutoff $\omega_c=\epsilon_F$ in the BCS gap equation for the repulsive term $\propto \ln \frac{\epsilon_F}{|\omega|}$ in Eq.~\eqref{eq:bilyr-Veff-small} introduces considerable net repulsive effective interactions. However these net repulsive effective interactions are absent in the full frequency dependent expression of Eq.~\eqref{eq:bilyr-Veff} and therefore unphysical.

\begin{figure}[ht]
  \centering
  \includegraphics[scale=0.60]{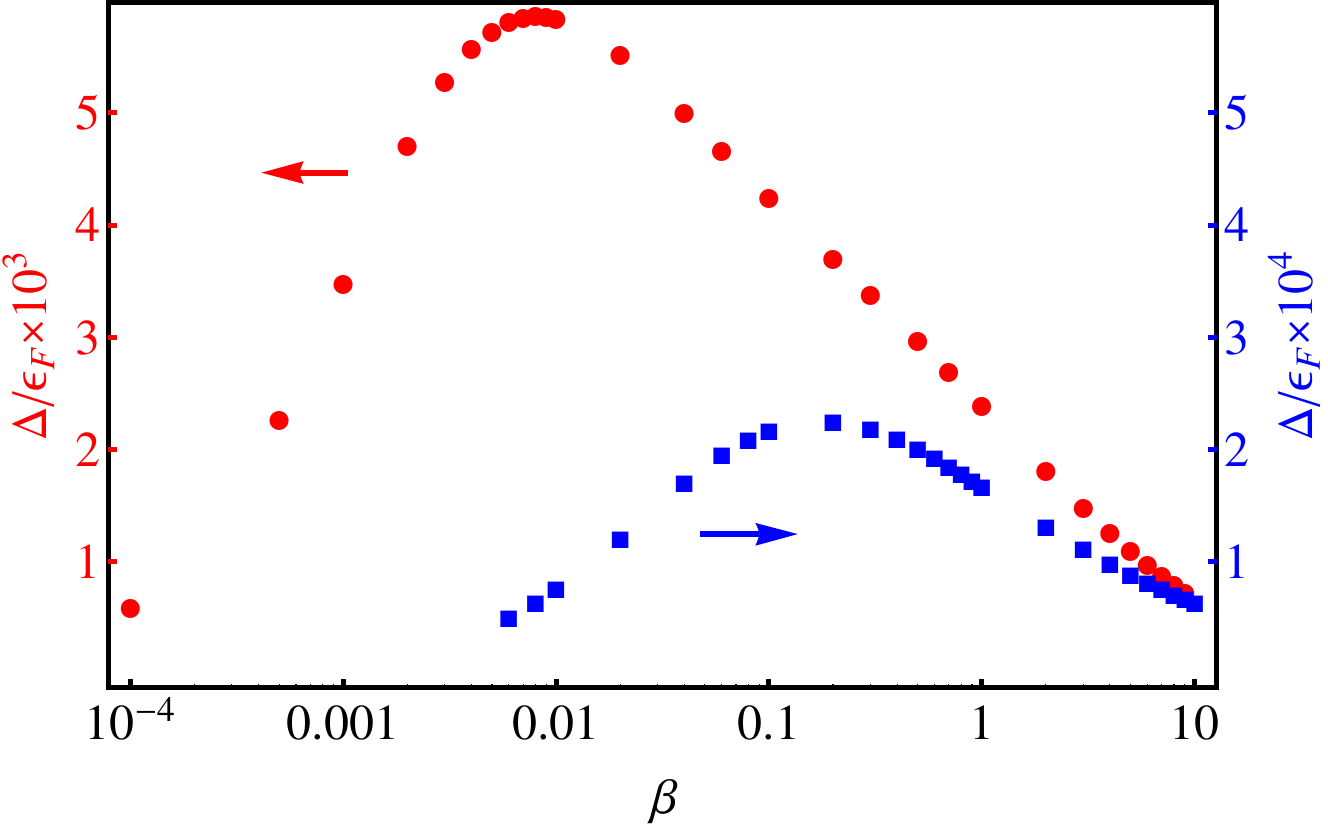}
  \caption{The bilayer pairing gap $\Delta$ as a function of the parameter $\beta$. The circles (in red) denote the data for $d=10\ell_B$; while the squares (in blue) denote the data for $d=50\ell_B$. Notice that the two plots are using two different vertical axis scales, as indicated by the two arrows.  Both sets of results were obtained using full frequency dependent $\widetilde{V}_{\rm{eff}}$, Eq.~\eqref{eq:bilyr-Veff} }
  \label{fig:bilyr-gap-beta-d-10-50}
\end{figure}

In Fig.~\ref{fig:bilayer-gap-beta-comparison} we also see that when $\beta$ is either very large or very small, the
gap tends to vanish. These two limiting cases can be understood as follows
\begin{itemize}
  \item When $\beta$ is very large, in the denominators of both $\mathcal{D}^{\pm}$, the Coulomb energy, dominating over the Landau damping, controls the gauge field fluctuations. Then in the effective interaction Eq.~\eqref{eq:bilyr-Veff} both terms are proportional to $\frac{1}{\beta}$, vanishing in the large $\beta$ limit. Therefore the net effective interaction is very small and the pairing gap tends to vanish in this limit.
  \item When $\beta $ is very small, the Landau damping term dominates over the Coulomb energy term and controls the gauge field fluctuations. However as Landau damping is the same for both the inter-layer in-phase and out-of-phase modes, they tend to cancel out each other in the effective interaction, as we can see in Eq.~\eqref{eq:bilyr-Veff}. Therefore the gap should vanish in this limit too.
\end{itemize}
From the considerations of these two limiting cases, we conclude that the gap must reach the maximum at some finite $\beta$ value. In fact this $\beta$ value depends on the specific value of the inter-layer distance $d$ as we see in Fig.~\ref{fig:bilyr-gap-beta-d-10-50}. 

\section{\label{conclusion}Conclusion}
To summarize, we have shown that the full frequency dependence of the effective interaction is  important and alters the conclusion obtained by Bonesteel\cite{Bonesteel1999}, which is based on a small frequency  analysis. For both short-ranged contact interaction  and  long-ranged Coulomb  interaction,  there can be a \emph{continuous} transitions (instead of a discontinuous one) from the HLR state to a chiral  pairing state in an odd angular momentum channel, as we tune the two coupling constants  which characterize the density-current and current-current interactions.  In practice, this tuning can be achieved by changing the width of the quantum well or of the semiconductor inversion layer, although the precise control of them can be difficult.

We have also constructed the phase diagrams for different  angular momentum channels and found that the phase diagram for $\ell=1$ channel pairing can be quite different from that for higher angular momentum channels $\ell \ge 3$. For $\ell=1$ we always need large enough density-current interaction to stabilize the chiral pairing state. But for $\ell \ge 3$,  even small density-current and large current-current interactions can result in a chiral pairing state, because of the attractive nature of the current-current interaction at high frequencies. As was pointed out in Ref.~\onlinecite{read2000}, for pairing in angular momentum channel $\ell$ there will be chiral Majorana fermion modes  on an edge, and correspondingly
$2^{|\ell|n-1}$degenerate states for 2n vortices. For $|\ell| =3$ pairing in $\nu=1/2$ state this will lead to nonabelian statistics.

We also applied the full frequency dependent analysis to the double layer half-filled problem considered previously by. Bonesteel \emph{et al.}~\cite{Bonesteel1996} It turns out that the full frequency dependence of the effective interaction does not change  their qualitative conclusion. This is because the two contributions to the effective interaction coming from the in-phase and out-of-phase mode fluctuations of the CS gauge field cancel each other out at high frequencies. Thus the net effective interaction is non-zero only in a very small frequency range near $\omega=0$. So the small frequency  analysis is a good approximation.~\cite{Bonesteel1996}

After our work was completed we noticed that   conclusions similar to ours  were recently reached by M. A. Metlitski et. al.~\cite{Metlitski2014}  from a renormalization group analysis. We also note two interesting related papers.~\cite{Cipri2014,Alicea2009} It is  clear that a more complex treatment of the coupled Eliashberg equations and fluctuation effects could  be subject of future work.
\begin{acknowledgments}
We thank N. Bonesteel and C. Nayak for a critical reading of an earlier version of the  manuscript and S. Raghu
for discussion.
S. C. and Z. W. were supported by US NSF under the Grant DMR-1004520. SC and IM were supported by the funds from the David S. Saxon Presidential Chair at UCLA. IM was also supported by  the Perimeter Institute for Theoretical Physics, and the John Templeton Foundation. Research at Perimeter Institute is supported by the Government of Canada through Industry Canada and by the Province of Ontario through the Ministry of Research and Innovation. S. B. C was funded by the Institute for Basic Science in Korea through the Young Scientist grant.
\end{acknowledgments}



%

\end{document}